\documentclass[twocolumn]{raa}
\usepackage{graphicx,times}
\usepackage{natbib}
\usepackage{amssymb,amsmath}
\bibpunct{(}{)}{;}{a}{}{,}

\usepackage[pagebackref=true]{hyperref}

\usepackage{hyperref}

\begin{document}
\title{Neutral Hydrogen content of dwarf galaxies in different environments}

\volnopage{Vol.0 (20xx) No.0, 000--000}      
\setcounter{page}{1}          

\author{Hui-Jie Hu\inst{1,3}, 
Qi Guo\inst{1,2,3}, 
Pablo Renard\inst{4}, 
Hang Yang\inst{1,3}
Zheng Zheng\inst{1,5}, 
Yingjie Jing\inst{1},
Hao Chen\inst{5}, 
and Hui Li\inst{4}
}


   \institute{National Astronomical Observatories, Chinese Academy of Sciences,
             Beijing 100012, China; {\it guoqi@nao.cas.cn, huhuijienao@gmail.com}\\
        \and
            Institute for Frontiers in Astronomy and Astrophysics, Beijing Normal University, Beijing 102206, China\\
        \and
            University of Chinese Academy of Sciences, No. 19 A Yuquan Road, Beijing 100049, China\\
        \and 
            Department of Astronomy, Tsinghua University, Beijing 100084, China\\
        \and
            Research Center for Intelligent Computing Platforms, Zhejiang Laboratory, Hangzhou 311100, China\\
\vs\no
   {\small Received 20xx month day; accepted 20xx month day}}

\abstract{Environments play an important role in galaxy formation and evolution, particularly in regulating the content of neutral gas. However, current H {\sc i} surveys have limitations in their depth, which prevents them from adequately studying low H {\sc i} content galaxies in high-density regions. In this study, we address this issue by employing the Five-hundred-meter Aperture Spherical radio Telescope (FAST) with extensive integration times to complement the relatively shallow Arecibo Legacy Fast Arecibo L-band Feed Array (ALFALFA) H {\sc i} survey. This approach allows us to explore the gas content of dwarf galaxies across various environments. 
We observe a positive relationship between H {\sc i} mass and stellar mass in dwarf galaxies, with a well-defined upper boundary for H {\sc i} mass that holds true in both observations and simulations. Furthermore, we find a decrease in the H {\sc i}-to-stellar mass ratio ($\rm M_{\rm HI}/M_*$) as the density of the environment increases, irrespective of whether it is determined by the proximity to the nearest group or the projected number density. Comparing our observations to simulations, we note a steeper slope in the relationship, indicating a gradual gas-stripping process in the observational data. Additionally, we find that the scaling relation between the $\rm M_{\rm HI}/M_*$ and optical properties can be improved by incorporating galaxy environments. 
\keywords{galaxies: dwarf --- galaxies: evolution --- galaxies: formation --- galaxies: statistics}
}

   \authorrunning{Hu. H et al.}            
   \titlerunning{H {\sc i} in different environments}  

   \maketitle

\section{Introduction}
The baryons in the Universe are primarily composed of hydrogen and helium, with hydrogen making up around 75\% of the total baryonic content. In the paradigm of the standard galaxy formation model \citep[e.g.][]{White1978,Springel2005,Guo2011} galaxies form when the gas cools, condenses, and forms stars at the centres of their dark matter haloes. When dark halos merge, their galaxies are accreted into more massive systems and orbit as satellite galaxies in groups and clusters. 

H {\sc i} gas, serving as the essential fuel for star formation, is indispensable for our understanding of galaxy formation and evolution. Different from the stellar component, H {\sc i} gas could be more susceptible to the environments as it can extend to larger distances compared to optical radii \citep{Haynes1984}. A significant portion of galaxies in the local Universe reside in groups and clusters \citep{Eke2006,Robotham2011}, where they suffer from various environmental influences such as tidal interactions, harassment, ram pressure stripping, and evaporation \citep[e.g.][]{Gunn1972,Moore1996,Denes2016,Cortese2021,Rhee2023}.

Over the last decades, numerous ground-based surveys: such as 2dF Galaxy Redshift Survey \citep[2dFGRS;][]{Colless2001}, the Sloan Digital Sky Survey \citep[SDSS;][]{York2000}, the United Kingdom Infrared Telescope (UKIRT) Infrared Deep Sky Survey \citep[UKIDSS;][]{Lawrence2007}, the Galaxy and Mass Assembly \citep[GAMA;][]{Driver2011}, the Two Micron All Sky Survey \citep[2MASS;][]{Skrutskie2006} Redshift Survey \citep[2MRS;][]{Huchra2012}, the Dark Energy Spectroscopic Instrument (DESI) surveys \citep{DESI2023,DESI2023b,Dey2019} and etc. and space-based surveys: and space-based: such as the Galaxy Evolution Explorer \citep[GALEX;][]{Martin2005}, the Wide-field Infrared Survey Explorer \citep[WISE;][]{Wright2010}, the Cosmic Evolution Survey \citep[COSMOS;][]{Scoville2007}, the Cosmic Assembly Near-infrared Deep Extragalactic Legacy Survey \citep[CANDELS;][]{Grogin2011}, and the ongoing James Webb Space Telescope (JWST) surveys have obtained vast amount of photometric and spectroscopic data for millions of galaxies in the local Universe and at high redshifts. These successful surveys have provided valuable insights into the evolution of stellar masses and star formation rates in galaxies (see the review paper and the reference in \citealt{Madau2014}). However, to fully understand the process of galaxy formation and evolution, it is crucial to understand the mechanisms of gas accretion onto galaxies and the efficiency with which that gas is converted into stars \citep[e.g.][]{Keres2005,Kennicutt1998,Salim2007,Leroy2008}. 

An economical approach to obtain the H {\sc i} is via galaxies' optical properties.  \citep{Kannappan2004} found that the H {\sc i}-to-stellar mass ratio, $\rm M_{\rm HI}/M_*$, exhibits a strong correlation with optical-optical (e.g., {\it u}-{\it r}) and optical-NIR (e.g., {\it u}-K) colors, allowing for its estimation with a typical scatter of approximately 0.4 dex. Subsequently, several studies have aimed to enhance the accuracy of photometric estimators for $M_{HI}/M_*$, by including stellar surface mass density, and optical \citep{Zhang2009} or near-ultraviolet (NUV)-optical \citep{Catinella2010,Zhangw2021} colors. These estimators typically yield a scatter of around 0.3 dex in log $\rm M_{HI}/M_*$, providing improved H {\sc i}-to-stellar mass ratio scaling relations.

Direct measurement of H {\sc i} relies on radio observations, which recently have also collected unprecedented data. For example, 
the blind H {\sc i} survey, the 100\% complete Arecibo Legacy Fast Arecibo L-band Feed Array \citep[ALFALFA;][$\alpha.100$]{Haynes2018} has detected approximately 30,000 extragalactic H {\sc i} line sources within a redshift of 0.06. However, due to the limited efficiency with an average integration time of 48 seconds for each source, this catalogue is rather shallow, with a flux limit of 0.18 Jy for a typical velocity width of 50 km/s, corresponds to 10$^{8.74} \rm M_{\odot}$ in H {\sc i} mass at z $\sim$ 0.0265. The Five-hundred-meter Aperture Spherical radio Telescope \citep[FAST;][]{Nan2006,Jiang2019}, with a 300-meter effective diameter and the ability to move within around 50 degrees, has the potential to significantly expand nearby H {\sc i} surveys. Similar to ALFALFA blind survey, using FAST Commensal Radio Astronomy FasT Survey \citep[CRAFTS;][]{Zhang2021} aims to detect nearly $4.8 \times 10^5$ galaxies up to a redshift of 0.1, covering a sky area of 20,000 deg$^2$. In this work, we use FAST to select 14 galaxies randomly from high-density and low-density environments and increase the integration time to around 19.2 minutes. This enables us to include gas-poor galaxies in various densities and shed light on how galaxies evolve in different environments.


In this study, we use a volume-limited sub-sample of the largest blind H {\sc i} survey, the 40\% complete ALFALFA \citep[$\alpha.40$;][]{Haynes2011} catalogue and gas-poor galaxies from FAST observation to examine the H {\sc i}-to-stellar mass ratio in different environments, and the scaling relations of H {\sc i}-to-stellar mass ratio with optical properties. In Section \ref{sec:data}, we briefly describe the sample selection criteria, the observation details, the data reduction, and the methods used to extract relevant physical properties. Our main results are presented and discussed in Section \ref{sec:res}. We summarize our results in Section \ref{sec:sum}.

\section{Data and Method} \label{sec:data}
Our aim is to investigate the H {\sc i}-to-stellar mass ratio in different environments. H {\sc i} fluxes are obtained using FAST and $\alpha.40$, while stellar mass is obtained using the data from the SDSS survey. 
\subsection{ALFLAFA dwarf galaxies} \label{sec:sele}
\begin{figure*}[ht!]
\includegraphics[width=17cm, angle=0]{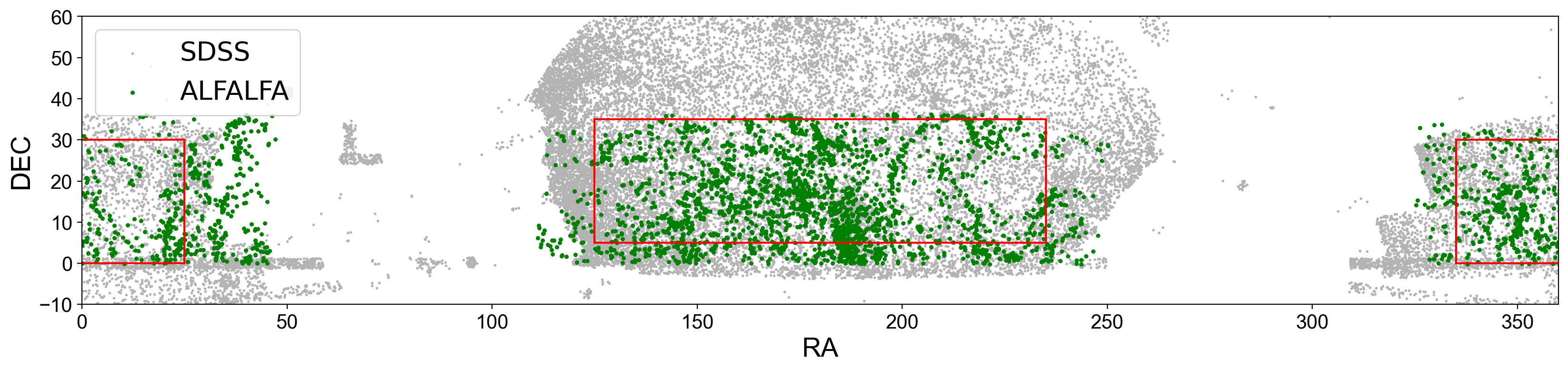}
\caption{{\bf The footprint of surveys.} The gray and green dots represent the SDSS spectroscopic galaxies within the redshift range of $0<z<0.017$ and ALFALFA galaxies within the redshift range of $0.003<z<0.0133$, respectively. The red rectangular regions represent the regions selected for the parent $\alpha.40$ dwarf galaxies. Further details can be found in the main text.
\label{fig:sdss}}
\end{figure*}

We select galaxies from the matched sample between the 40\% complete Arecibo Legacy Fast Arecibo L-band Feed Array \citep[ALFALFA, a.k.a. $\alpha.40$;][]{Haynes2011} survey and Seventh Data Releases of SDSS \citep[SDSS DR7;][]{Abazajian2009}. 

The SDSS uses a 2.5 m telescope located at Apache Point Observatory (APO) near Sacramento Peak in Southern New Mexico. The main goal is to obtain photometry in five broad bands over 10,000 deg$^2$ of high-latitude sky, and spectroscopy of a million galaxies and 100,000 quasars over this same region. SDSS DR7 comprises images and spectra for millions of galaxies, centered at a redshift $\sim$ 0.1. The extinction-corrected spectroscopic data are complete at 17.72 mag in {\it r}-band.

In previous studies, it was discovered that the SDSS photometric pipeline tends to overestimate the sky background of galaxies with extended low surface brightness by $\sim$ 0.5 mag \citep{He2013,Liu2008,Lisker2007}. This is particularly relevant for dwarf galaxies with low surface brightness as their luminosities are more vulnerable to sky subtraction. To address this issue, we selected a sub-sample of low surface brightness galaxies \citep[LSBGs;][]{Du2015} in $\alpha.40$, with central surface brightness in the B band $\mu_{0}$(B) $>22.5$ mag arcsec$^{-2}$, and recalculated their photometry using the {\it g}-band and {\it r}-band SDSS mosaic images from the Twelfth Data Releases of the Sloan Digital Sky Survey \citep[SDSS DR12;][]{Alam2015}, implementing improvements in sky-background subtraction following \cite{Zheng2015} and \cite{Guo2020}. For galaxies with brighter surface brightness, we utilized Petrosian magnitudes from the released SDSS DR7 catalogue. 
 
The stellar mass is estimated using the stellar mass-to-light ratio formula proposed by \cite{Bell2003}
\begin{equation}
\log_{10}({M/L}) = a_{\lambda}+b_{\lambda} \times (g - r)
\end{equation}
Here we adopt the Kroupa stellar initial mass function \citep{Kroupa2002} with $a_r = -0.306-0.15$ and $b_r = 1.097$, $a_i = -0.222-0.15$ and $b_i = 0.864$ \citep{Bell2003}.

The ALFALFA survey is the biggest blind extragalactic H {\sc i} survey, with a wide sky coverage of 7,000 deg$^2$. The spatial resolution beam size is about 3.5 arcmin. The catalogue encompasses 21-cm line spectra of more than 30,000 extragalactic sources with radial velocities $<$18,000 km s$^{-1}$. The average integration time of ALFALFA sources is 48s, corresponding to a minimum flux of 0.18 Jy at a velocity width of 50 km/s. Here we use the 40\% complete ALFALFA, a.k.a. $\alpha.40$. There are 15,855 H {\sc i} detected sources in $\alpha.40$, among which 15,041 are extragalactic objects and the rest 814 are possibly to be Galactic high-velocity clouds. Cross-matched with the SDSS DR7, there are 12,423 galaxies with optical counterparts \citep{Haynes2011}. 

\begin{figure}[ht!]
\includegraphics[width=8.5cm, angle=0]{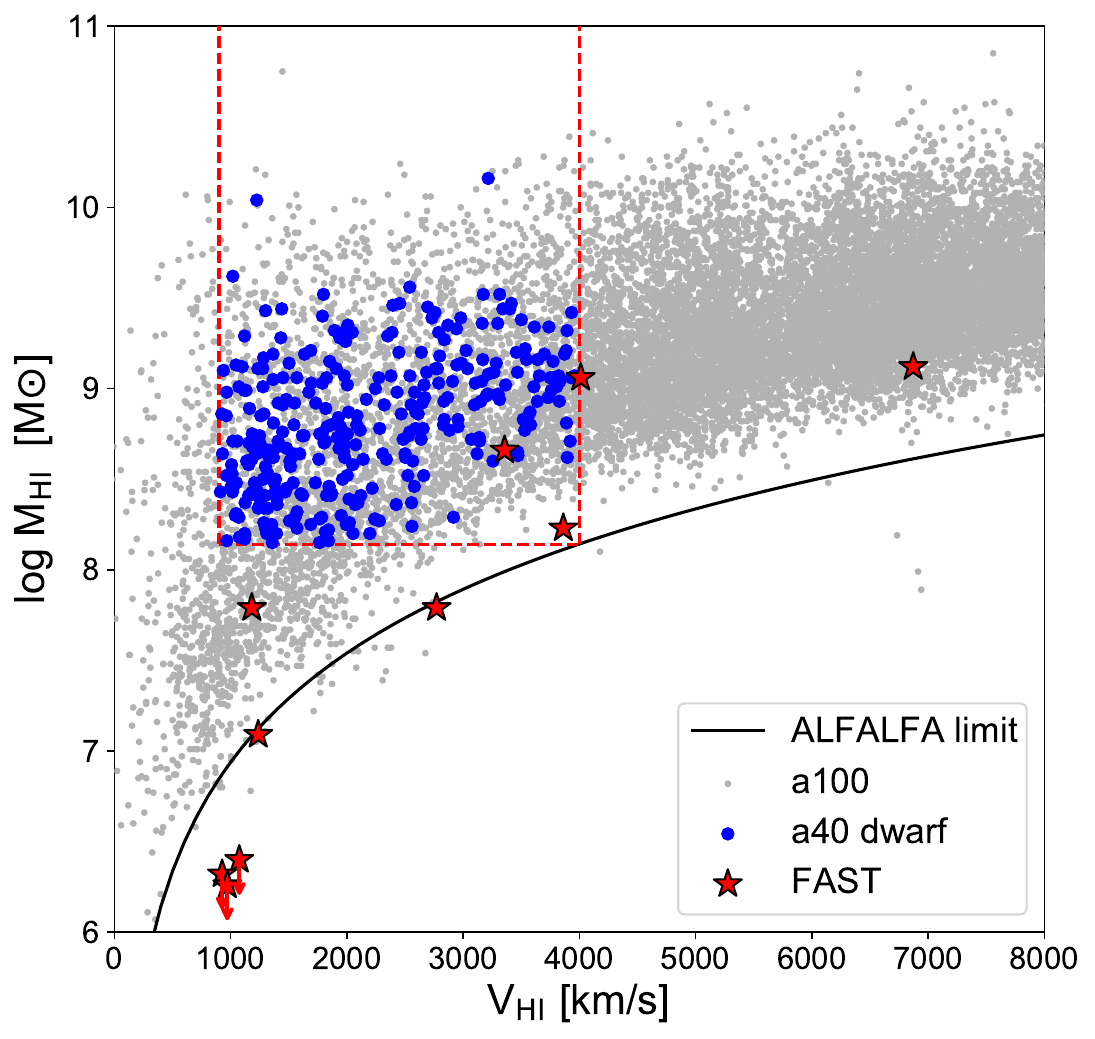}
\caption{{\bf H {\sc i} mass v.s. redshift.} Gray dots represent galaxies from the $\alpha.100$ sample, while the blue dots depict our volume-limited dwarf sample selected from $\alpha.40$. Red stars correspond to the galaxies observed by FAST. The arrows denote the three FAST targets which have an upper limit of $\rm M_{HI}<10^7M_{\odot}$. Red dashed lines denote the selection boundaries. The black curve illustrates the detection limits of H {\sc i} mass, assuming a velocity width of 50 km/s. 
\label{fig:comp}}
\end{figure}

We select dwarf galaxies following the criteria in \cite{Hu2023}. Firstly, we select galaxies with absolute magnitude $\rm M_r > -18$ and the H {\sc i} spectra with high signal-to-noise ratios, S/R$_{HI} > 10$. Then, we visually remove the galaxies which have multiple optical counterparts within $6'$ (corresponding to twice the Arecibo beam size), which leaves 770 dwarf galaxies. Secondly, we use the regions where both SDSS and ALFALFA have detection, as shown in Fig. \ref{fig:sdss}. In order to avoid the incompleteness of neighbor galaxies at the edges of the survey, we further restrict to $5^{\circ}<DEC<35^{\circ}$, $125^{\circ}<RA<235^{\circ}$; and $0^{\circ}<DEC<30^{\circ}$, $-25^{\circ}<RA<25^{\circ}$, and we request the samples to be within a redshift range (0.003, 0.0133), leading to 413 selected galaxies. 
Finally, we restrict our sample to have {\it r}-band magnitude brighter than 20 and H {\sc i} mass $> 10^{8.14} M_{\odot}$. This {\it r}-band limit is selected to have an optically complete sample set at 95\% level \footnote{From COMBO S11 field, see on SDSS website: https://www.sdss4.org/dr17/imaging/other\_info}, and this H {\sc i} mass limit is chosen to ensure an H {\sc i} mass complete sample in the redshift of interest (Fig. \ref{fig:comp}). The final sample consists of 320 dwarf galaxies. 

\subsection{Deep observations with FAST}
The ALFALFA survey has provided a substantial sample of galaxies with detectable H {\sc i} content. However, due to the relatively short integration time of the survey, the sample mainly consists of H {\sc i}-rich dwarf galaxies. In order to gain a comprehensive understanding of the variability of H {\sc i} content in different environments, including H {\sc i}-poor galaxies, we have increased the integration time and utilized the Five-hundred-meter Aperture Spherical Telescope \citep[FAST;][]{Nan2006,Jiang2019} to observe a set of 10 randomly selected dwarf galaxies, which have angular distances to any $\alpha.100$ sources greater than $6'$ (twice the Arecibo beam size). These galaxies either have not been detected in the ALFALFA survey or have low signal-to-noise ratios (S/N, as shown in Fig. \ref{fig:comp}). By expanding our observations to include these targets, we aim to obtain a more comprehensive understanding of the distribution of H {\sc i} content across various environments. 

FAST is a single-dish radio telescope with an effective diameter of 300 meters and the highest sensitivity among ground-based single-dish radio telescopes. FAST is currently mounted with a 19-beam receiver covering a frequency range of 1.05–1.45GHz. Its frequency resolution is about 7.63 kHz (65,536 channels over 1.0–1.5 GHz bandwidth) and the angular resolution is about 2.9 arcmin at low redshift for 21-cm line spectra. The system temperature is $\sim$20K for zenith angles within 26.4 deg observations. 

The observations were carried out based on a `Shared-Risk' project, using the position switch ON-OFF mode recording two polarizations (XX and YY) for each target. The ON-source and OFF-source integration time is 120 seconds in most cases, 
and the overhead time (switch between ON-source and OFF-source positions) is 30 seconds. We have carefully chosen the OFF point for each source so that a side beam (M08 or M14\footnote{We found that some delta RFI only appeared in the M14 beam so we only used the M14 beam for those M08 positions with contamination}) would be pointing to the source when the central beam (M01) is pointing to the OFF point \citep{Zheng2020}. This strategy successfully reduced the overhead time rates down to around 20\%. For most of our targets, the sampling time is 0.1 s, we inject the high (10K) noise diode signal for 0.1 s every 1 s during the observation. For AGC 125644, we use a 0.5 s sampling time and inject the high noise diode signal for 2 s every 20 s. The majority of the observations have a total on-source integration time of approximately 19.2 minutes. Details are listed in Table \ref{tab1}. 

We perform the following procedures for each beam and their corresponding polarizations (XX and YY).

{\bf Flux calibration.} We use the high (10K) noise diode other than the standard calibrators for our flux calibrations. First, we convert the recorded noise diode signal into antenna temperature (K). To obtain the flux density in units of Janskys (Jy), we utilized the gain values reported in Table 5 of \cite{Jiang2020}, for the M01 and M08 (M14) beams, respectively. 

{\bf Baseline and standing wave removal.} Following \cite{Jing2023}\footnote{HiFAST: https://hifast.readthedocs.io}, we remove a lower-order Polynomial baseline to each OFF-subtracted spectrum (ON-source subtract OFF-source for each beam, i.e. M01 ON-source position minus M01 OFF-source position \footnote{Initially, our plan was to subtract the continuum using the same time from OFF position, specifically the M01 ON-source position minus the M08 ON-source position (pointing the OFF-source position). However, we discovered that the period and phase of the standing wave differed between the different beams, making it challenging to fit the standing wave if we performed the same time OFF subtraction. We noticed that the standing wave varies with time, but we were able to subtract the ripple of the spectra by performing the same beam OFF subtraction.}). There are $\sim$ 1 MHz standing waves covering all bandwidth of FAST spectra, which are caused by reflections between the receiver and the bottom panel of the telescope (138.63 meters causing a 1.08 MHz standing wave). We remove the standing wave by fitting a sinusoidal function using spectra around the heliocentric H {\sc i} velocity of the target. 

The observations were conducted in topocentric mode, which means that the reference frame is not a rest frame due to the Earth's movement. To account for this, we convert the observed frequency to heliocentric velocity. The final spectra are obtained by averaging the spectra from M01 and M08 (M14). 

\begin{figure*}[ht!]
\includegraphics[width=17cm, angle=0]{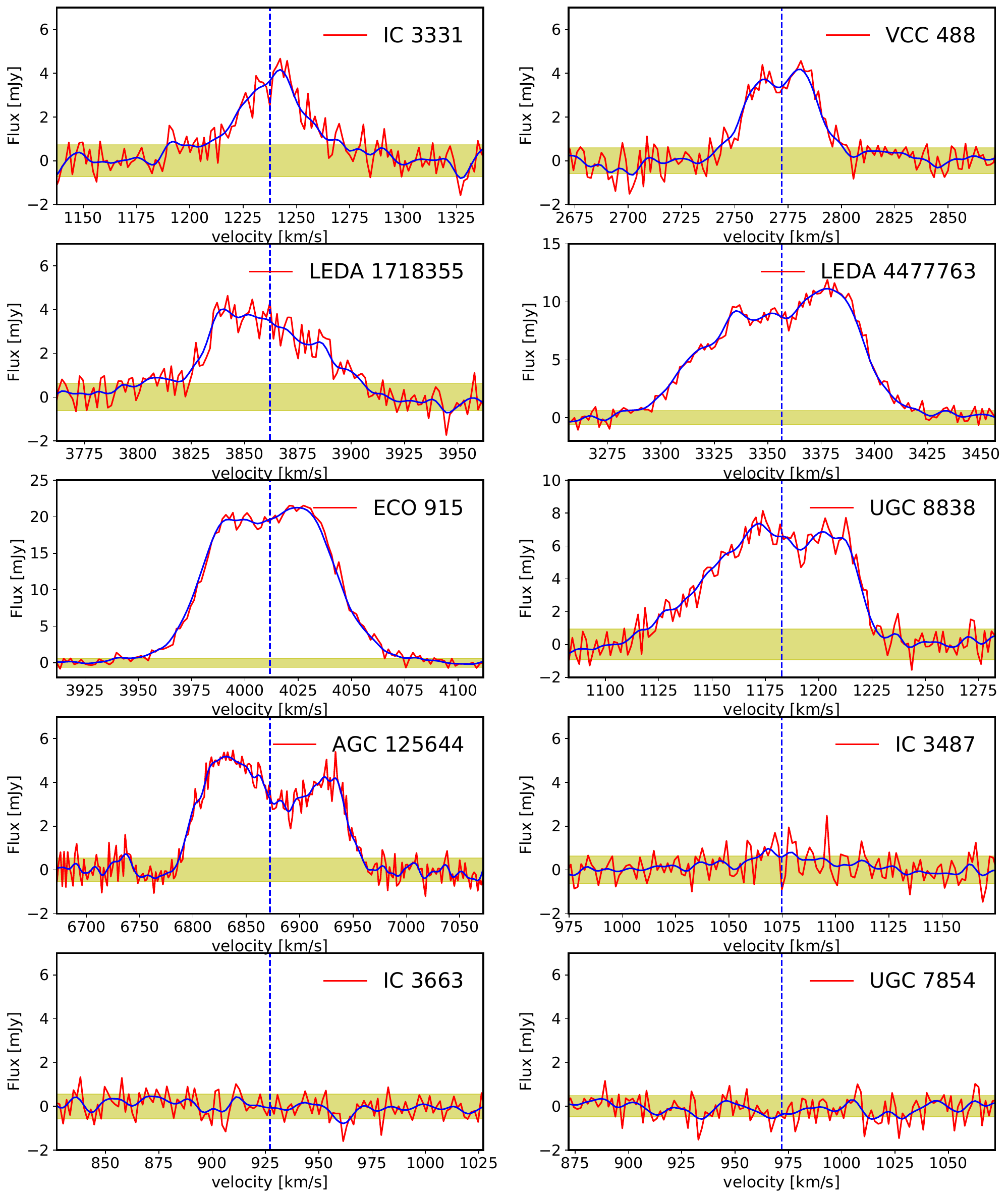}
\caption{{\bf H {\sc i} spectra of FAST targets.} Each panel shows the spectrum of an individual galaxy detected by FAST using a long integration time (Table \ref{tab1}). Red curves are the FAST observations with a frequency resolution of 7.63 kHz ($\sim 1.65$ km/s), and blue curves depict the spectra smoothed using a Hanning smoothing window. The yellow regions indicate the 1$\sigma_{rms}$ regions of baselines. Blue dashed lines represent their heliocentric velocity (center velocity of $W_{50}$). 
\label{fig:spec}}
\end{figure*}

The spectra of detected FAST sources are presented in Fig. \ref{fig:spec}. Out of these sources, 7 of them have good signal-to-noise ratios (S/N), which allows for reliable analysis of their H {\sc i} properties. However, for the remaining 3 sources that have low S/N, their spectra only provide upper limits for the analysis. These upper limits are used to estimate the potential H {\sc i} content in these sources and we incorporate them into the overall analysis. 

\begin{figure}[ht!]
\includegraphics[width=8.5cm, angle=0]{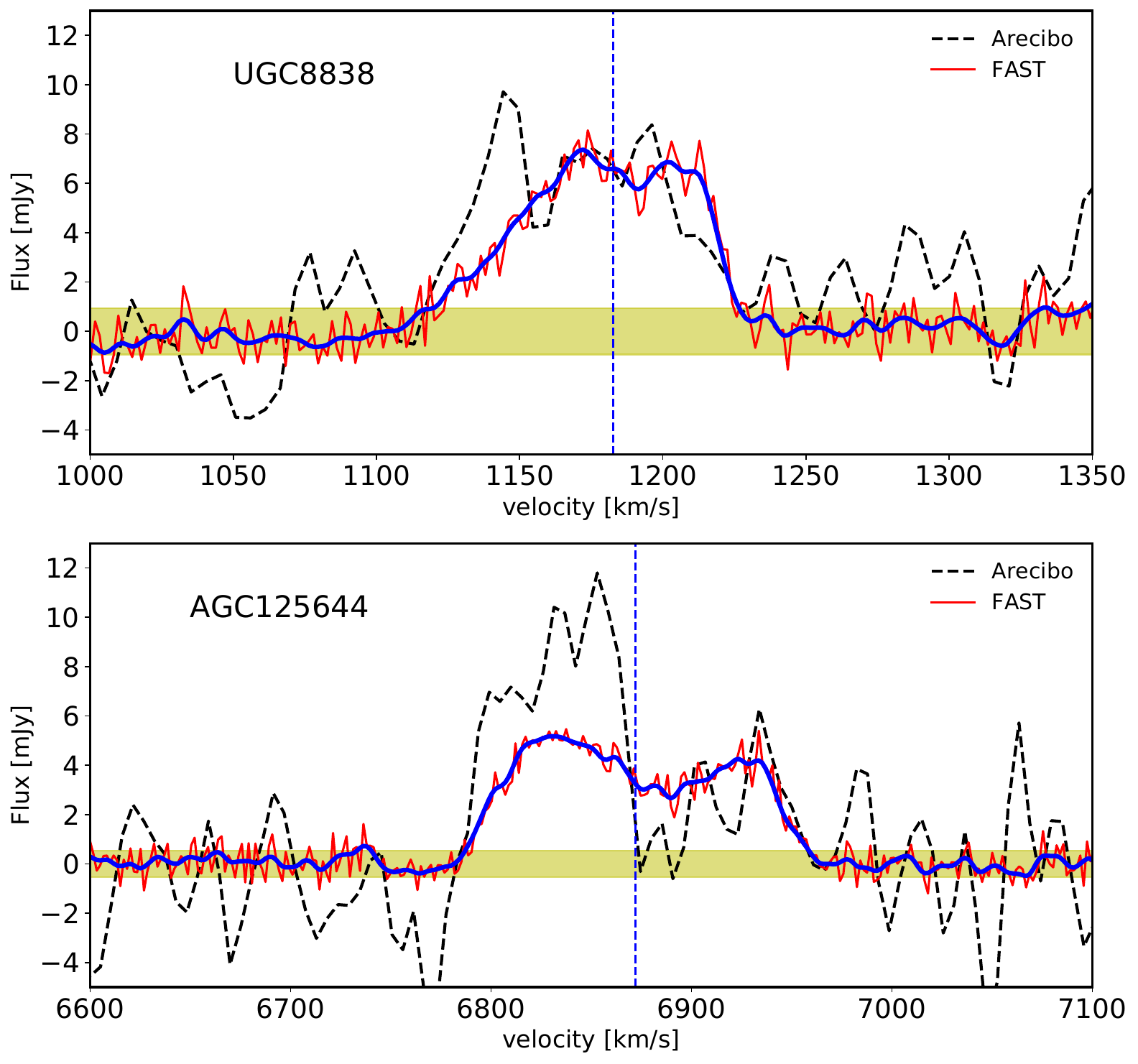}
\caption{{\bf Comparison of H {\sc i} spectra between deep FAST observation and $\alpha.40$.} 
The blue dashed lines correspond to their heliocentric velocity (center velocity of $W_{50}$). The black curves depict the spectra from ALFALFA, while the red curves are the FAST observations with a frequency resolution of 7.63 kHz ($\sim 1.65$ km/s). The solid blue curves depict the FAST spectra smoothed using a Hanning smoothing window. The yellow regions indicate the 1$\sigma_{rms}$ regions of baselines. The upper panel displays the H {\sc i} spectrum of UGC 8838, while the lower panel shows the spectrum of AGC 125644. 
\label{fig:specalf}}
\end{figure}

To illustrate the advantage of using longer integration time with the FAST telescope for H {\sc i}, we choose two galaxies for comparison from the ALFALFA sample: UGC 8838 and AGC 125644. The spectral profile of UGC 8838 in Figure \ref{fig:specalf} exhibits a relatively low signal-to-noise ratio (S/N), while AGC 125644 displays a dubious sub-component between 6900 km/s and 7000 km/s, also with a very low S/N. Figure \ref{fig:specalf} illustrates that, by utilizing a longer integration time, FAST effectively improved the signal-to-noise ratio (S/N) of UGC 8838 from 7.5 to 14.8. For AGC 125644, we confirmed the existence of the suspicious sub-component, despite the S/N in ALFALFA already being as high as 7.1. This discovery led to an expansion of its line width by a factor of two compared to its measurement in ALFALFA. Overall, it demonstrates the improved detection capabilities of the FAST telescope through longer integration times, enabling the identification of previously unidentified sub-components and enhancing the S/N ratio of the observed galaxies.

We future extract the physical properties as follows, with the derived properties listed in Table \ref{tab2}. 

{\bf Velocity width $\bf W_{\bf 50}$ and $\bf W_{\bf 20}$}. The algorithm follows the procedure in \cite{Guo2020}. We distinguish between systems exhibiting one peak and two peaks in their H {\sc i} spectra. For sources with a single peak, we identify the wavelengths at which the fluxes are 50\% (20\%) of the peak values to obtain the raw values, $W_{50,raw}$ ($W_{20,raw}$). For sources with two peaks, the 50\% (20\%) fluxes are calculated for each side, and we identify the corresponding characteristic wavelengths at each rising side of the spectrum to measure the velocity width. We perform polynomial fitting at each rising side \citep{Springob2005,Haynes2011} to obtain the characteristic wavelengths. 
We then subtract the instrumental broadening as follows \citep{Catinella2012}:
\begin{equation}
    W_{50(20)} = \sqrt{W_{raw}^2-\Delta v^2}
\end{equation}
where $\Delta v$ represents the velocity resolution, which is approximately 1.65 km s$^{-1}$. 

{\bf The heliocentric velocity and H {\sc i} line integrated flux density}, $S_{21}$, in units of Jy km s$^{-1}$. The heliocentric velocity is defined as the center velocity of $W_{50}$. The total flux is then computed by integrating the continuum subtracted H {\sc i} spectrum within the $\pm 2\times W_{50}$ range from the heliocentric velocity. The 5 $\sigma$ upper limits of the flux density for three undetected galaxies (IC 3487, IC 3663, and UGC 7854) are calculated as $5\sqrt{W_{50}/1.65 \times \sigma_{rms}^2} \times 1.65$ under the assumption that $W_{50}=100$ km/s. 

{\bf Signal to noise ratio (S/N) of the spectrum.} We perform two different estimations of S/N: 

(1) An S/N estimation analog to ALFALFA following \cite{Haynes2011}, the S/N of the detection is estimated as:
\begin{equation}
    S/N = (S_{21}/W_{50}) \times \omega_{smo}^{1/2}/\sigma_{rms}
\end{equation}
where $\omega_{smo}$ is a smoothing width set to be $W_{50}/20$ for $W_{50}< $ 400 km s$^{-1}$, and to be 20 for $W_{50}\ge $ 400 km s$^{-1}$, and $\sigma_{rms}$ is the root-mean-square of the baseline.

(2) An average S/N within velocity width $W_{50}$: 
\begin{equation}
    (S/N)_{avg} = (S_{21}/W_{50})/\sigma_{rms}
\end{equation}

{\bf H {\sc i} mass and distance.} In ALFALFA, the distances for galaxies with $\rm cz>6000$ km s$^{-1}$ are the Hubble distances, estimated as $\rm d = cz/H_0$ where $H_{0}$=70 km s$^{-1}$ Mpc$^{-1}$ is the Hubble constant, while for galaxies with $cz<6000$ km s$^{-1}$, the distances are obtained with a combination of primary distances from the literature and secondary from the TFR. For our newly observed FAST targets, we only use the primary distances in the literature, as they are more accurate. When no accurate distance measurements are available in the literature, we use Hubble distance instead which is calculated from the heliocentric velocity of the H {\sc i} spectrum. The H {\sc i} mass of the galaxy is then derived from the integrated H {\sc i} flux and distance following \cite{Roberts1962}, under the assumption that the H {\sc i} content is optically thin:
\begin{equation}
{M_{HI}} = 2.36\times 10^{5}d^{2}\int {S_{21}} \mathrm{d} {v [M_{\odot}]}
\end{equation}
where d is the luminosity distance to the galaxy unit in Mpc. 


\subsection{Dwarf galaxies in TNG50}
We make use of the Illustris-TNG cosmological hydrodynamical simulations\footnote{https://www.tng-project.org} for the comparison. TNG50 \citep{Nelson2019,Pillepich2019} traces $2 \times 2160^3$ dark matter particles and gas cells in a period box size of 51.7 Mpc. The mass resolution is $4.5 \times 10^5M_\odot$ and $8.5 \times 10^4M_\odot$ for dark matter and baryon, respectively. Stellar masses are calculated using all star particles within each subhalo. We adopt the ``Neutral Hydrogen Abundance" from the catalogue for H {\sc i} mass in non-star-forming cells and calculate the cold hydrogen gas mass (H {\sc i} $+\rm H_2$) using the two-phase interstellar medium model (a modification of \citealt{Springel2003}) in star-forming cells. Then, we drop out the molecular hydrogen from all cold hydrogen mass to obtain the H {\sc i} mass for star-forming cells. The molecular hydrogen fraction $f_{H_2}$ in star-forming regions is calculated by the modified KMT model \citep{Krumholz2009,McKee2010} as

\begin{equation}
    f_{H_2} = \left\{
     \begin{array}{lr}
         1 - 3s/(4+s)\   &  \rm{if}\ s < 2 \\
         0\     & \      \rm{if}\ s \geq 2.
     \end{array}
    \right.
\end{equation}
where {\it s} is given by
\begin{equation}
s = \frac{\ln{(1 + 0.6\chi + 0.01\chi^2)}}{0.6\tau_c}
\end{equation}
and
\begin{equation}
\chi = 0.756(1+3.1Z^{0.365}) 
\end{equation}

\begin{equation}
\tau_c = \frac{\Sigma_{\rm{gas}}\sigma}{\mu}
\end{equation}
where $Z$ represents the metallicity for gas cell in units of solar metallicity $Z_{\odot} = 0.0127$, $\sigma = Z \times 10^{-25}\  \rm{m^2}$ is the dust cross-section and $\mu = 3.9\times10^{-27}\ \rm{kg}$ is the mean mass per particles. The Jeans length approximation method is adopted when estimating the gas surface density $\Sigma_{gas}$. We discard galaxies with less than 200 star particles to ensure a reliable estimation of stellar mass. 

\begin{figure}[ht!]
\includegraphics[width=8.5cm, angle=0]{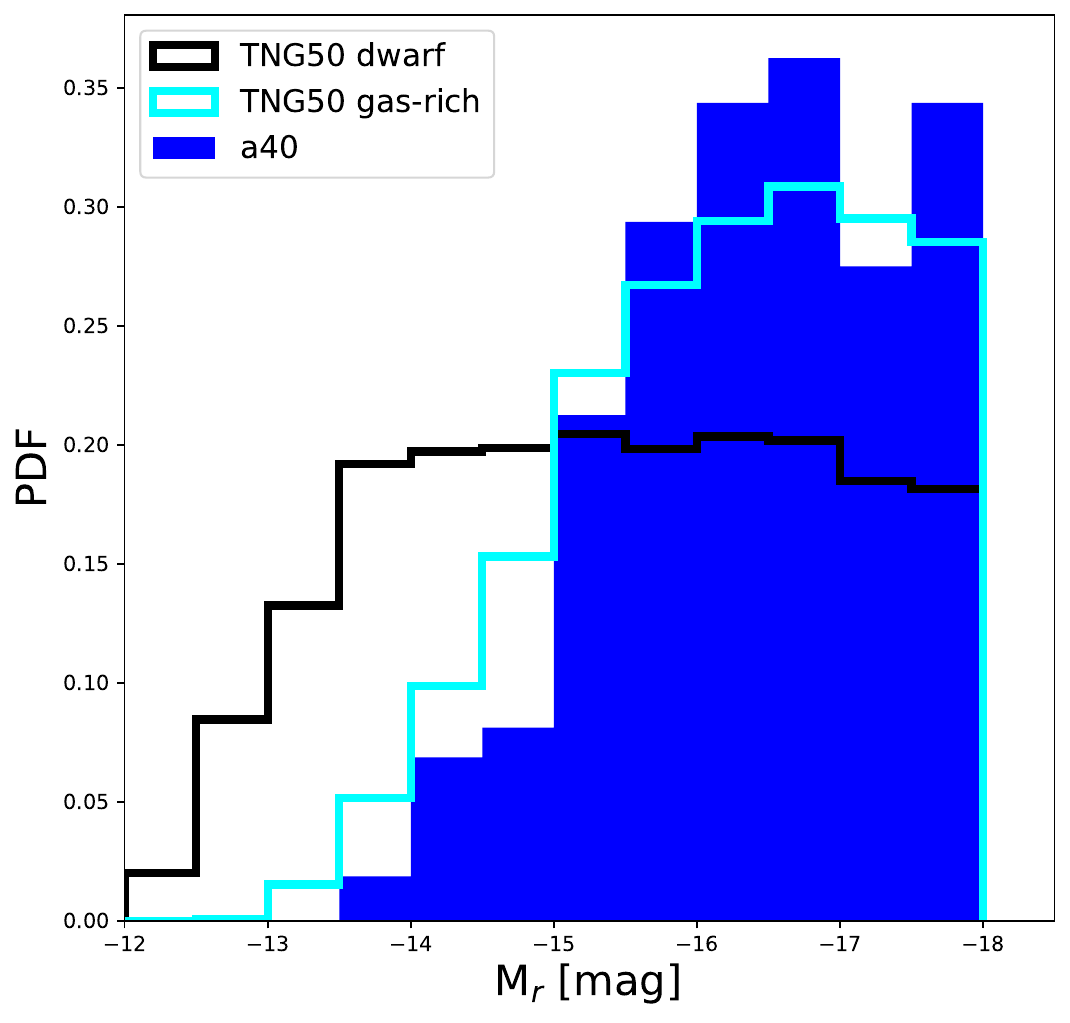}
\caption{{\bf Absolute {\it r}-band magnitude distribution.} The blue-filled histogram represents the parent $\alpha.40$ dwarf galaxies, while the black and cyan histograms correspond to the TNG50 dwarf and TNG50 gas-rich dwarf galaxies. 
\label{fig:mr}}
\end{figure}

In order to compare to observations,  here we select dwarf galaxies with $\rm M_r > -18$ and H {\sc i} mass $\rm > 10^{8.14} M_{\odot}$ for the main comparison (TNG50 gas-rich dwarf galaxies). While we did not set a lower limit for the {\it r}-band luminosity, Figure \ref{fig:mr} demonstrates that the distributions in $\rm M_r$ are similar between the $\alpha.40$ dwarf sample and the TNG50 gas-rich dwarf galaxy sample. Furthermore, to examine the impact of environmental effects, we utilized a larger sample by including all galaxies with $\rm M_r > -18$, which encompasses both gas-rich and gas-poor galaxies. As depicted in Figure \ref{fig:mr}, it is evident that such selection also encompasses a broader range of faint galaxies in terms of {\it r}-band magnitude.

\subsection{Environment}
\label{env}
\begin{figure}[ht!]
\includegraphics[width=8.5cm, angle=0]{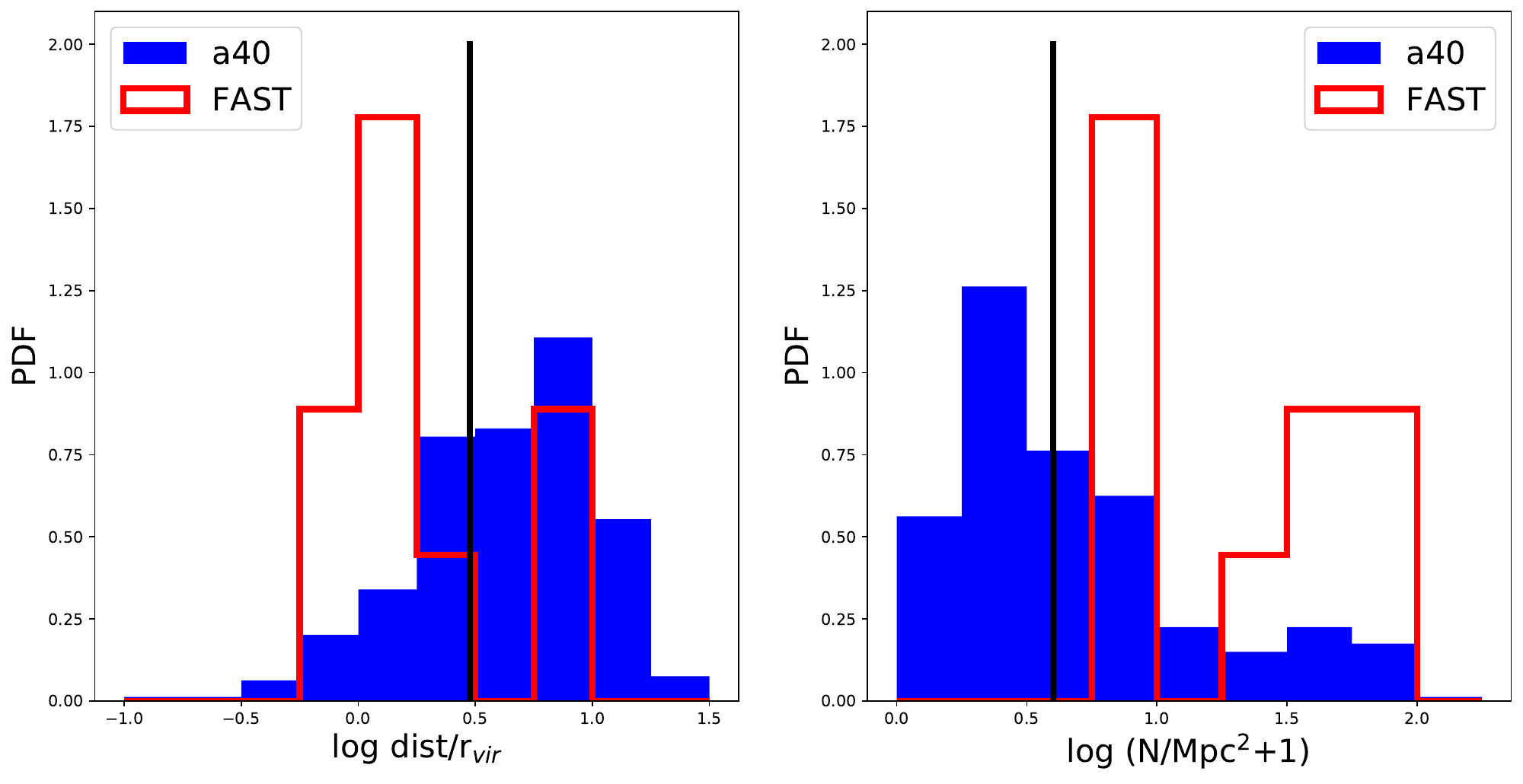}
\caption{{\bf Environment distributions.} The left panel shows the histogram of the distance to the nearest group, while the right panel displays the histogram of the projected number density. The blue histograms represent the parent $\alpha.40$ dwarf galaxies, while the red histograms correspond to the FAST targets. The black vertical solid lines indicate three times the virial radius (left panel) and the median number density value (right panel) of the $\alpha.40$ dwarf galaxies. 
\label{fig:envpdf}}
\end{figure}

We adopt two kinds of environment definitions, the projected number density and the distance to the nearest group/cluster.

The projected number density is defined as the number of galaxies brighter than -16.6 mag ({\it r}-band) within 1 Mpc projected distance and up to $\pm 1000$ km/s velocity difference along the line-of-sight. Here we use the spectroscopic data of SDSS which is complete at 17.72 mag in {\it r}-band, corresponding to an absolute magnitude of $\sim -16.6$ mag at the redshift of 0.017 (z=0.0133$+ \frac{1000 km/s}{c}$). 

Distance to the nearest group is defined as the minimum value of the distance to the group center in a unit of the corresponding group viral radius \citep{Guo2020}. The group catalogue is taken from  \citep{Saulder2016} which applies a friends-of-friends (FoF) group finder algorithm on the SDSS DR12 \citep{Alam2015} and the 2MASS Redshift Survey \citep[2MRS;][]{Huchra2012}. Compared to the group catalogue by \cite{Yang2007}, the group catalogue used in this paper includes more groups in the local Universe (including groups at $z<0.01$). Various observational biases have been taken into account, including the Malmquist bias, the `Fingers of God', etc. The groups/clusters masses are derived from the total luminosity, luminosity distance, velocity dispersion, group radius, and the number of detected group members (we refer the reader to \citealt{Saulder2016} for more details). This method has been precisely calibrated on the mock catalogues from the Millennium simulation \citep{Guo2011,Guo2013}. The minimum and median distances to the nearest group of the parent dwarf galaxies are 0.074 and 4.339, respectively. 

The histograms of two different definitions of the environment are shown in Fig. \ref{fig:envpdf}. Most $\alpha.40$ dwarf galaxies reside in relatively low-density regions, outside the virial radius of surrounding clusters/groups. Galaxies at high densities or close to groups/clusters suffer from environmental stripping and likely lose their H {\sc i} gas in a short time scale. One needs a longer integration time to reach deep detection of galaxies in such environments. FAST, with its longer integration time, captures a higher fraction of galaxies in these high-density regions, as depicted by the red histogram. Accordingly, the combination of the gas-rich $\alpha.40$ galaxies and the dwarf galaxies residing in denser regions, exposed for a longer duration through FAST, form a valuable sample for investigating environmental dependencies.


\begin{table*}
\bc
\caption{{\bf Observational information.} Columns: (1) Galaxy name, (2) \& (3) coordinate, (4) spectroscopic redshift, (5) distance, (6) observed date, (7) integration time over total observing time unit in minutes, (8) FAST beam numbers which were used for integration, (9) comments. Most of our targets are observed during 2019 (PID: 2019a-003S), while AGC 125644 is observed using the PID: ZD2021\_4.\label{tab1}}
\large
 \begin{tabular}{ccccccccccccc}
  \hline\noalign{\smallskip}
Name & RA & DEC & z & distance & date & $t_{int}/t_{tot}$ & Beam & comment \\
 & degree[$^{\circ}$] & degree[$^{\circ}$] &  & [Mpc] &  & min/min & \\
  \hline\noalign{\smallskip}
IC 3331 & 186.52215 & 11.81221 & 0.00411 & 17.6 & 6/30/2019 & 16/24 & 1, 14 & \\
VCC 488 & 185.41364 & 7.25363 & 0.00921 & 39.5 & 6/30/2019 & 19.2/24 & 1, 8 & \\
IC 3487 & 188.30599 & 9.39735 & 0.00359 & 16.2 & 6/30/2019 & 19.2/24 & 1, 8 & No signal\\
IC 3663 & 190.41421 & 12.24749 & 0.00309 & 15.7 & 7/18/2019 & 19.2/24 & 1, 8 & No signal \\
UGC 7854 & 190.46653 & 9.40295 & 0.00324 & 15.9 & 7/18/2019 & 19.2/24 & 1, 8 & No signal \\
LEDA 1718355 & 175.34892 & 24.85794 & 0.01280 & 54.9 & 8/19/2019 & 19.2/24 & 1, 8 & \\
LEDA 4477763 & 236.80355 & 17.95755 & 0.01110 & 47.6 & 8/19/2019 & 19.2/24 & 1, 14 &  \\
ECO 915 & 237.41397 & 18.88038 & 0.01328 & 56.9 & 8/19/2019 & 19.2/24 & 1, 8 &  \\
UGC 8838 & 208.91125 & 4.98472 & 0.00390 & 22.2 & 8/19/2019 & 8/24 & 1 &  \\
AGC 125644 & 33.305 & 4.63583 & 0.02278 & 93.8 & 1/11/2022 & 15.7/20 & 1, 8 & PID: ZD2021\_4 \\
  \noalign{\smallskip}\hline
\end{tabular}
\ec
\end{table*}

\begin{table*}
\bc
\caption{{\bf FAST sources parameters.} Columns: (1) Galaxy name, (2) \& (3) velocity widths, (4) heliocentric velocity, (5) stellar mass (from {\it r}-band estimation) (6) neutral hydrogen mass, (7) RMS under 1.65/km velocity resolution (8) \& (9) signal-to-noise ratios, (10) distance to the nearest group environment, (11) projected number density environment. \label{tab2}}
\large
 \begin{tabular}{cccccccccccccc}
  \hline\noalign{\smallskip}
Name & $W_{50}$ & $W_{20}$ & $V_{helio}$ & log $M_{*}$ & log $M_{HI}$ & RMS & S/N & S/N & Env & Env \\
 & [km/s] & [km/s] & [km/s] & [$M_{\odot}$] & [$M_{\odot}$] & [mJy] & average & ALFALFA  & group & N \\
  \hline\noalign{\smallskip}
IC 3331 & 29.2 & 49.93 & 1237.6 & 8.65 & 7.09 & 0.73 & 7.9 & 9.6 & 1.03 & 56 \\
VCC 488 & 34.43 & 45.52 & 2772.0 & 8.35 & 7.79 & 0.59 & 8.3 & 10.9 & 7.56 & 19 \\
IC 3487 &  &  & 1074.8 & 8.34 & 6.40$\downarrow$ & 0.66 &  &  & 0.93 & 68 \\
IC 3663 &  &  & 927.1 & 8.55 & 6.32$\downarrow$ & 0.63 &  &  & 1.76 & 43 \\
UGC 7854 &  &  & 971.9 & 8.79 & 6.26$\downarrow$ & 0.51 &  &  & 1.57 & 41 \\
LEDA 1718355 & 56.19 & 60.75 & 3861.9 & 8.57 & 8.23 & 0.62 & 6.9 & 11.5 & 3.16 & 6 \\
LEDA 4477763 & 79.07 & 106.72 & 3356.6 & 8.16 & 8.66 & 0.61 & 17.6 & 35 & 7.95 & 5 \\
ECO 915 & 67.83 & 89.41 & 4011.7 & 8.42 & 9.06 & 0.61 & 35.9 & 66.1 & 1.37 & 6 \\
UGC 8838 & 73.07 & 98.99 & 1182.7 & 9.04 & 7.79 & 0.94 & 7.7 & 14.8 & 0.64 & 7 \\
AGC 125644 & 139.97 & 166.79 & 6872.1 & 8.88 & 9.12 & 0.55 & 8.2 & 21.8 & 8.56 &  \\
  \noalign{\smallskip}\hline
\end{tabular}
\ec
\end{table*}


\section{Results} 
\label{sec:res}
In this section, we investigate the relationship between the H {\sc i} and stellar mass, their environmental dependence, and the scaling relations between the H {\sc i}-to-stellar mass ratio and optical properties.

\subsection{H {\sc i} to stellar mass relation}
\begin{figure*}[ht!]
\includegraphics[width=17cm, angle=0]{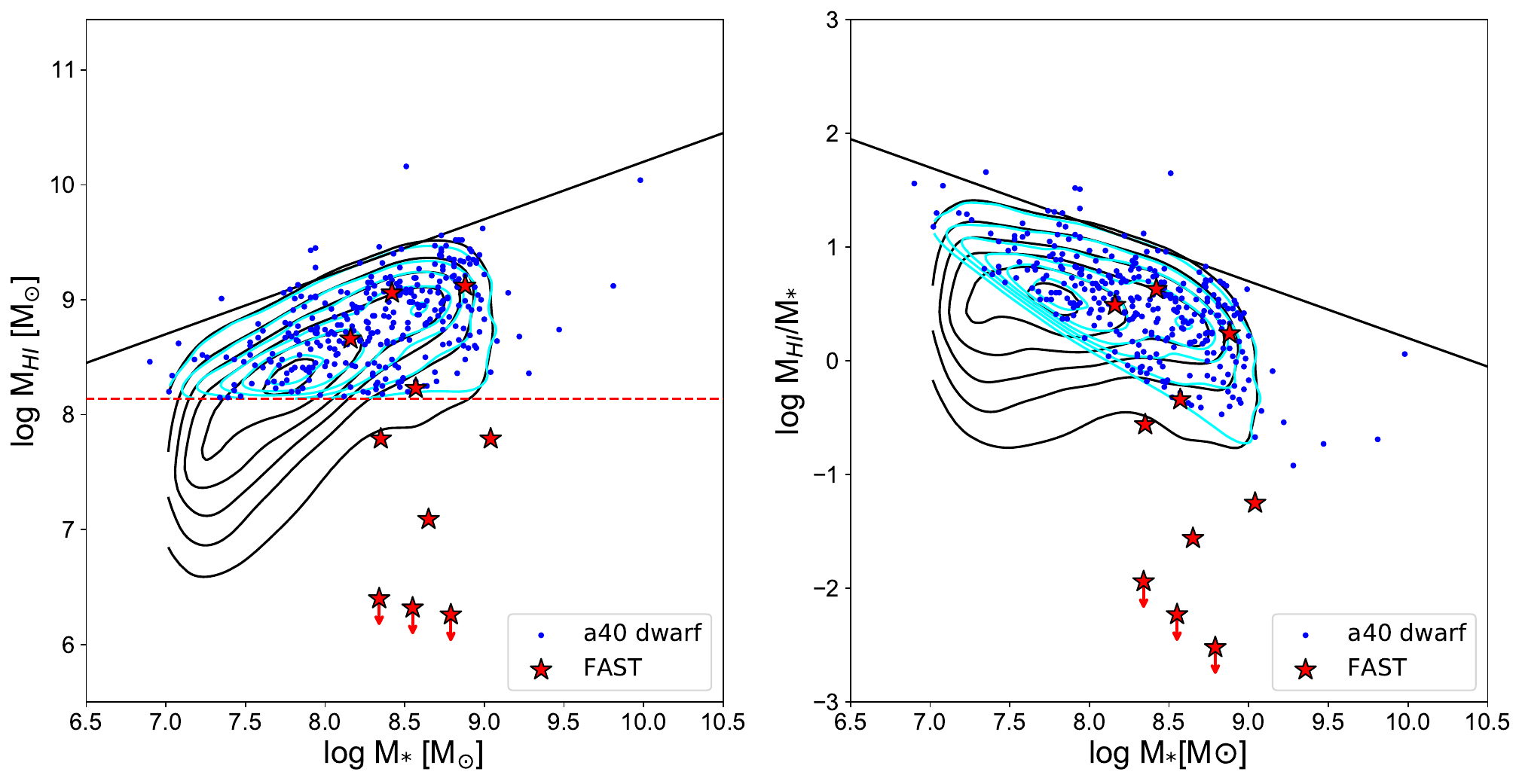}
\caption{{\bf Left:} H {\sc i} mass v.s. stellar mass relation. The red dashed line represents the completeness limit of $\rm M_{HI}=10^{8.14}M_{\odot}$. {\bf Right:} the H {\sc i}-to-stellar mass ratio v.s. stellar mass relation. The black and cyan contours correspond to the TNG50 dwarf and TNG50 gas-rich dwarf samples, respectively. The blue dots represent parent $\alpha.40$ dwarf galaxies, while the red stars indicate the FAST targets. The black lines depict the upper limit of the contours for the TNG50 sample. 
\label{fig:smHI}}
\end{figure*}

Figure \ref{fig:smHI} presents the relationship between H {\sc i} mass and stellar mass. The H {\sc i} mass of the jointly selected gas-rich dwarf galaxy sample appears to increase with their stellar masses. The observed H {\sc i} v.s. stellar mass relation is well reproduced in simulations when taking into account the detection limit of the $\alpha.40$. However, the gas-poor galaxies detected by FAST push this H {\sc i} mass v.s. stellar mass relation downwards more than an order.  

When including both gas-rich and gas-poor dwarf galaxies in simulations, it shows an even stronger dependence of the H {\sc i} mass with stellar masses. Dwarf galaxies with low H {\sc i} masses as detected by FAST are also found in simulations. Indeed, TNG50 exhibits a continuous distribution along the y-axis, with a relatively small fraction of galaxies having $\rm log M_{HI} < 7.5$ at $\rm log M_*\sim 8.5$.

Interestingly, we observe a common boundary for the maximum $\rm M_{HI}$ (H {\sc i} mass) in both our observational data and simulations. This boundary sets a limit on the highest H {\sc i} mass that can exist for a given stellar mass ($\rm M_*$). Such behavior also translates into the $\rm M_{HI}$ to $\rm M_*$ ratio as a function of stellar mass (right panel). To describe this upper boundary, we employed a fitting function as follows. 
\begin{equation}
\log_{10}({M_{HI}/M_{*}}) = -0.5 \times \log_{10}({M_{*}})+5.2
\end{equation}
This is equivalent to 
\begin{equation}
 \frac{M_{HI}^2}{M_{*}} = 10^{10.4}M_{\odot}
\end{equation}
Given the detection limit, at least 97.2\% $\alpha.40$ dwarf galaxies are below this boundary, while in simulations for TNG50 dwarf galaxies, 99.2\% are below this boundary. 

\subsection{H {\sc i}-to-stellar mass ratio in different environments}
\begin{figure*}[ht!]
\includegraphics[width=17cm, angle=0]{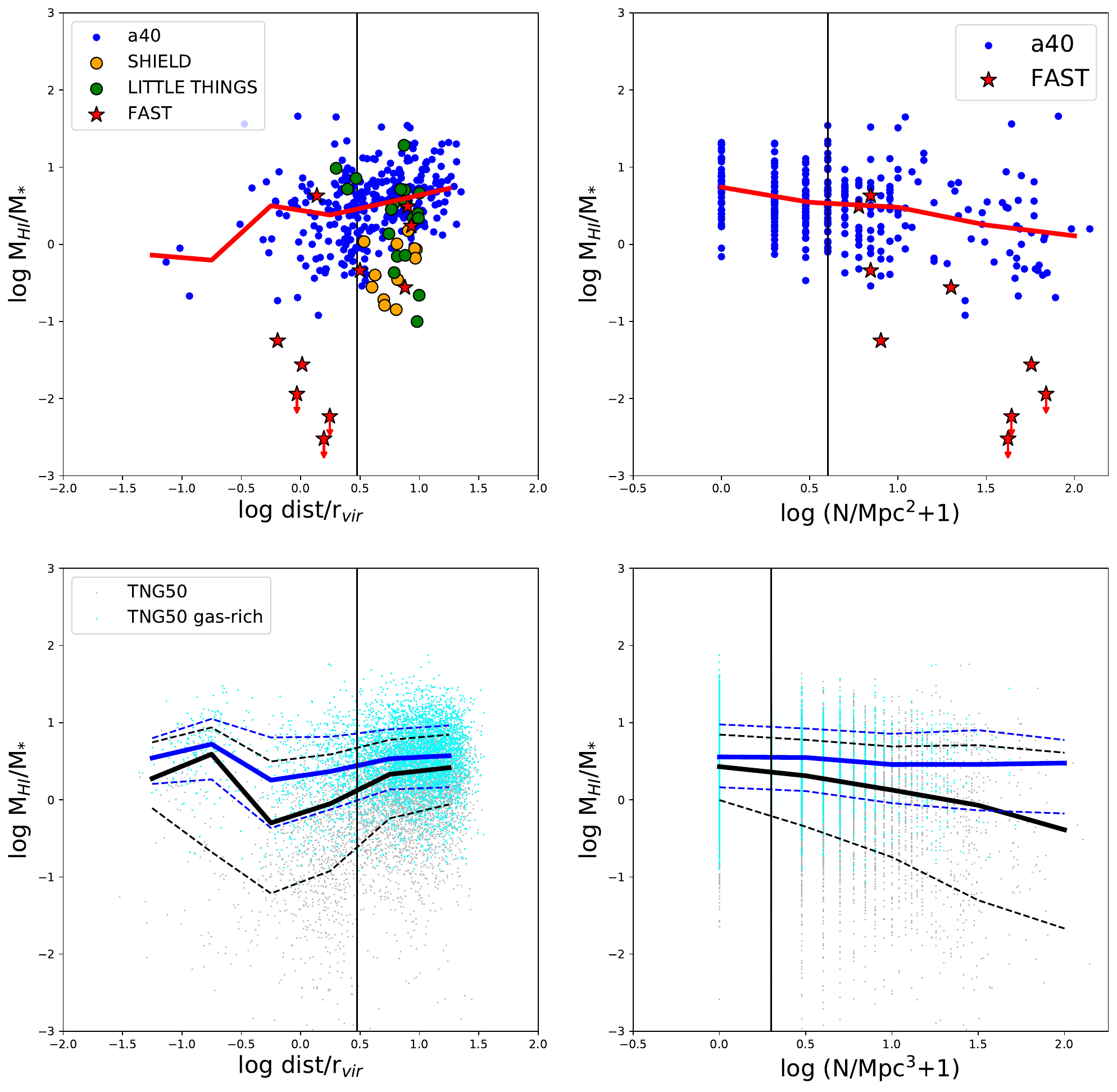}
\caption{{\bf H {\sc i}-to-stellar mass ratio as a function of environments. Upper left panel:} Results for the observational data. The blue dots are the $\alpha.40$ dwarf galaxies, with the red line denoting the median $\rm M_{\rm HI}/M_*$ value. The orange and the green dots are dwarf galaxies from SHIELD and LITTLE THINGS. The red stars are our FAST targets, and the arrows indicate the upper limit of undetected sources. The black line indicates the critical radius of three times the virial radius. The environment is defined by the distance to the nearest group. {\bf Upper right panel:} Similar to the upper left panel but the environment is defined by projected number density. The black line represents the median number density. {\bf Bottom left panel:} similar to the upper left panel but for the TNG50 galaxies. The gray and cyan dots represent the TNG50 dwarf and TNG50 gas-rich dwarf galaxies. The dark and blue curves represent the corresponding 16\%, 50\%, and 84\% percentage levels. {\bf Bottom right panel:} Similar to the upper right panels but for the TNG50 galaxies. Data have the same line style and color coding as those in the bottom left panel.}
\label{fig:HIFlog}
\end{figure*}

Figure \ref{fig:HIFlog} displays the distribution of the H {\sc i}-to-stellar mass ratio ($\rm M_{\rm HI}/M_*$) across different environments. To extend the sample size, we include data from two Local Volume dwarf galaxy samples: Local Irregulars That Trace Luminosity Extremes, The H {\sc i} Nearby Galaxy Survey \citep[LITTLE THINGS;][]{Oh2015} and Survey of H {\sc i} in Extremely Low-mass Dwarfs \citep[SHIELD;][]{McNichols2016}. 

\begin{figure}[ht!]
\includegraphics[width=8cm, angle=0]{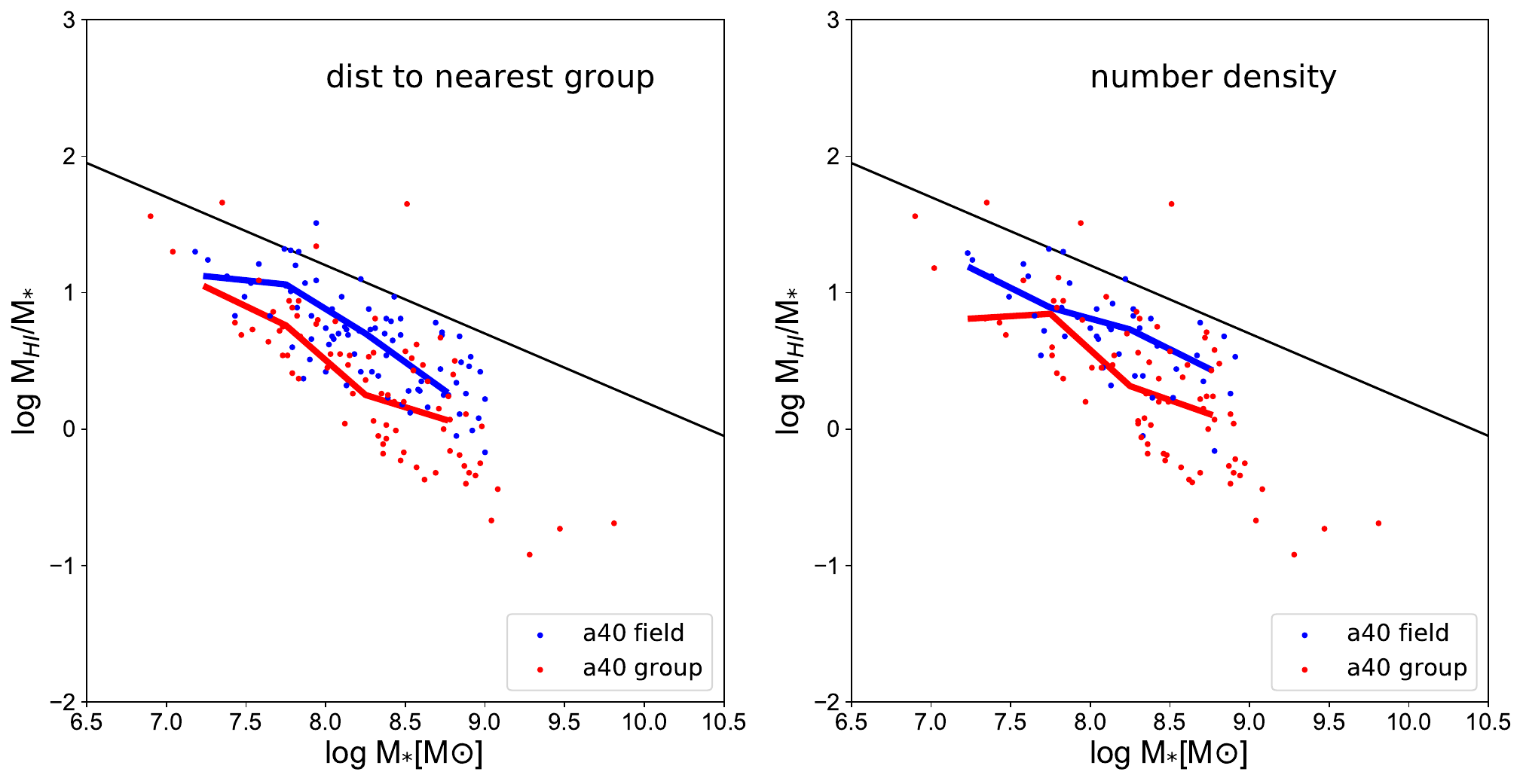}
\caption{{\bf H {\sc i}-to-stellar mass ratio as a function of stellar mass in different environments. Left panel:} The environment is defined by the distance to the nearest group. {\bf Right panel:} The environment is defined by the number density. Blue dots are the $\alpha.40$ dwarf galaxies in the 25\% most isolated environments, while red dots are galaxies in the 25\% most dense environments.
\label{fig:HIFenv}}
\end{figure}

The left panels of the figure demonstrate a slight decrease in the $\rm M_{\rm HI}/M_*$ as the distance to the nearest group decreases. The black lines in the left panels indicate where the distance to the nearest group is three times the group virial radius, beyond which the high-density environmental effects typically vanish. In the right panels, the black lines correspond to the median values of the number density for the $\alpha.40$ dwarf galaxies (3) and TNG50 dwarf (1). A consistent trend is observed, indicating a decrease in the $\rm M_{\rm HI}/M_*$ as the environment becomes denser, regardless of the specific definition of the environment. To remove the mass dependence, Fig. \ref{fig:HIFenv} shows the $\rm M_{\rm HI}/M_*$ vs. $\rm M_*$ relations for galaxies in the 25\% most dense
 and the 25\% most isolated environments. It is clear that galaxies in dense regions have a lower $\rm M_{\rm HI}/M_*$ at any given stellar mass. 

We observe a lower boundary of the $\rm M_{\rm HI}/M_*$ ratio at around 0.1, which could be an artifact resulting from the limited observation depth of the $\alpha.40$ survey. At low densities galaxies detected by FAST reside in the same region as the $\alpha.40$ sample. However, the high sensitivity of FAST, along with longer integration times, extends the lower boundary by a factor of 10 near groups/clusters. These findings suggest that there is a gradual and continuous process of H {\sc i} gas stripping around groups/clusters, rather than a sudden removal. 

Such environmental dependence is much weaker, if any, in simulations. In the case of TNG50 gas-rich dwarf galaxies which mimic the observed $\alpha.40$ dwarf sample, their $\rm M_{\rm HI}/M_*$ ratio does not vary much when galaxies get closer to groups/clusters, or when the density increases. This suggests that gas stripping occurs faster in simulations than those in the real Universe. When including gas-poor dwarf galaxies in TNG50, it encloses the regions covered by the gas-poor dwarf galaxies detected through the deep FAST observation. The dependence on environments is stronger compared to the TNG50 gas-rich sample, but overall, the dependence remains weak.  

In summary, the process of gas stripping from galaxies as they approach high-density regions is rapid, but the timescale of this process could be even shorter in simulations than those in observations.

\subsection{H {\sc i}-to-stellar mass ratio v.s. galaxy optical properties}
\begin{figure*}[ht!]
\includegraphics[width=17cm, angle=0]{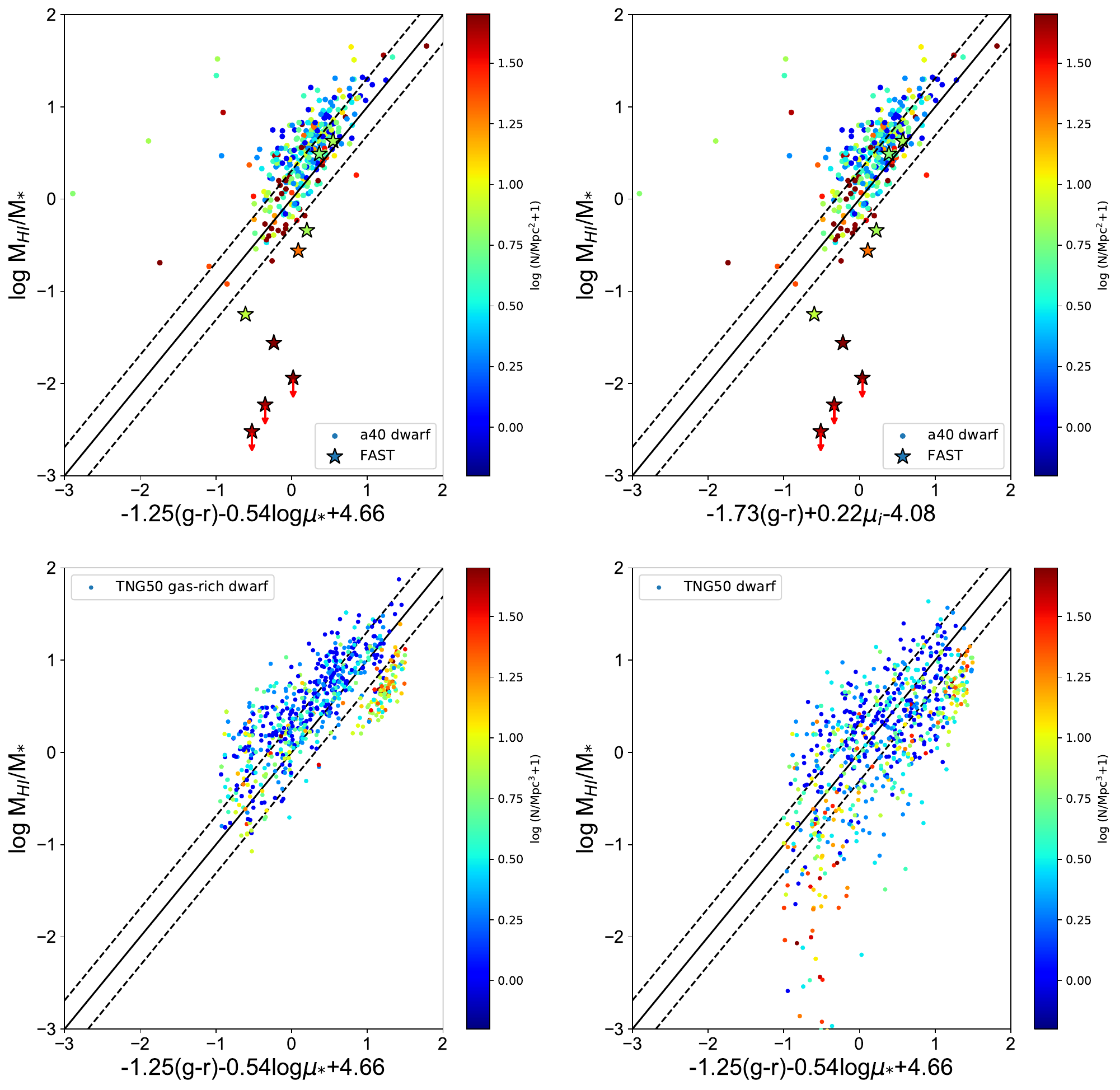}
\caption{{\bf Scaling relations between H {\sc i} to stellar mass ratio and optical properties. Upper left panel:}$\rm M_{\rm HI}/M_*$ v.s. $f(g-r, \mu_*)$ The dots and stars represent the $\alpha.40$ dwarf galaxies and the FAST targets. Symbols are colored by log(number density $+$1). {\bf Upper right panel:} Similar to the upper left panel but for $\rm M_{\rm HI}/M_*$ v.s. $f(g-r, \mu_i)$. {\bf Bottom left panel:} Similar to the upper left panel but for TNG50 gas-rich dwarf galaxies. {\bf Bottom right panel:} Similar to the upper left panel but for TNG50 dwarf galaxies. The black solid lines and dashed lines indicate the scaling relation and its 1-$\sigma$ scatter established by \cite{Zhang2009} 
\label{fig:scal}}
\end{figure*}

Previous studies found that H {\sc i}-to-stellar mass ratio is closely related to galaxies' optical properties \citep{Zhang2009,Catinella2010,Li2012,Zhangw2021}. 
In Figure \ref{fig:scal}, we present the relation between $\rm M_{\rm HI}/M_*$ and optical proprieties, as proposed by \cite{Zhang2009}. In the upper left panel, we show the $\rm M_{\rm HI}/M_*$ as a function of the combination of the {\it g}-{\it r} color and stellar surface mass density. The stellar surface mass density is defined as $\rm log(\mu_{*,i}) = log(M_{*}) - log(2\pi R^2_{50,i})$, where $\rm M_{*}$ represents the total stellar mass, and $\rm R_{50,i}$ corresponds to the {\it i}-band Petrosian half-light radius. In the upper right panel, the stellar surface mass density is replaced with surface brightness, which is defined as $\rm \mu_i = m_i + 2.5 log(2\pi R^2_{50,i})$, where $m_i$ represents the apparent Petrosian {\it i}-band magnitude. \cite{Zhang2009} used a sample of 800 galaxies from the SDSS DR4 and the HyperLeda H {\sc i} \citep{Paturel2003} catalogue (Sample II) and obtain the scaling relations as presented by the solid lines, with the dashed lines corresponding to the 1$\sigma$ scatter.

Figure \ref{fig:scal} demonstrates that the H {\sc i}-selected sample from $\alpha.40$ exhibits a slightly higher $\rm M_{\rm HI}/M_*$ compared to the sample studied by \cite{Zhang2009}. This divergence could be attributed to the exclusion of gas-poor galaxies in $\alpha.40$, which are primarily found in high-density regions. To address this, we incorporate the inclusion of high-density gas-poor galaxies detected by the FAST telescope, particularly in high-density environments. When analyzing the samples separately based on different environments, a clear dependence on the environment becomes apparent. For a given combination of the optical properties, galaxies in dense regions seems to have lower $\rm M_{\rm HI}/M_*$.

In the lower panels of Fig. \ref{fig:scal}, we present a comparison with the TNG50 gas-rich dwarf (left) and TNG50 dwarf galaxies (right). For gas-rich dwarf galaxies, it shows a dependence on environments mainly on the right-hand side where the gas fraction is relatively high. When including gas-poor dwarf galaxies, $\rm M_{\rm HI}/M_*$ reaches lower values at the left-hand side, galaxies in dense regions are presented well below the scaling relations found by \cite{Zhang2009}.  

These findings imply that the current understanding of the relationship between the $\rm M_{\rm HI}/M_*$ and optical properties may not be accurate enough without considering the influence of environmental factors. 
 

\section{Summary} \label{sec:sum}
In this paper, we investigate the H {\sc i}-to-stellar mass ratio ($\rm M_{\rm HI}/M_*$), the scaling relations between $\rm M_{\rm HI}/M_*$ and optical properties, and their dependence on environments using a jointly selected volume-limited sample from $\alpha.40$ survey and SDSS. Our sample primarily consists of gas-rich dwarf galaxies, and to complement this dataset with more gas-poor galaxies, we incorporate data from 10 dwarf galaxies observed with the FAST telescope. These additional observations were conducted with significantly longer integration times, allowing for a more comprehensive analysis. Our main findings are as follows:

1) There exists a clear correlation between H {\sc i} mass and stellar mass in our jointly selected sample of dwarf galaxies. This finding is also supported by simulations. Furthermore, we observe a common upper boundary for H {\sc i} mass among both the simulation and observational datasets for each given stellar mass. This suggests that there is a limit to the amount of H {\sc i} mass that dwarf galaxies can possess, regardless of their stellar mass.

2) The H {\sc i}-to-stellar mass ratio exhibits a decreasing trend as the environment becomes denser, regardless of whether it is defined by the distance to the nearest group or the (projected) number density. These findings are further supported by our analysis of deeper observations of high-density gas-poor targets. Simulations also exhibit similar trends, but the slope is much shallower. This suggests that the stripping of gas in high-density regions occurs more gradually in the real Universe compared to simulations. 

3) The $\alpha.40$ dwarf galaxies show a slight departure from the median of the $\rm M_{HI}/M_*$ v.s. optical properties scaling relation established by \cite{Zhang2009}, with a tendency towards higher $\rm M_{HI}/M_*$ values while still mostly falling within the one-sigma region.
We also observe a dependency on the environment for both the observed and simulated datasets, although this effect is less pronounced in the simulations.

The majority of our findings are derived from the $\alpha.40$ dwarf galaxies. Although we have added a gas-poor sample with 10 dwarf galaxies through deep observations with FAST, we acknowledge that the statistical significance remains limited. Therefore, a more comprehensive and extensive survey is necessary to determine if these galaxies are indeed outliers or if they simply contribute to the scatter observed in the scaling relations.

\begin{acknowledgements}
This work is supported by the National SKA Program of China (No. 2022SKA0110201 and 2022SKA0110200), the National Key Research and Development of China (No.2018YFA0404503), and CAS Project for Young Scientists in Basic Research Grant No. YSBR-062, the National Natural Science Foundation of China (NSFC) grants (No.12033008, 11622325, 11988101), the K.C.Wong Education Foundation, and the science research grants from the China Manned Space Project (CMSP) with NO.CMS-CSST-2021-A03 and NO.CMS-CSST-2021-A07. P.R. is supported by the National Science Foundation of China (grant no. 12073014), the science research grants from the China Manned Space Project with No. CMS-CSST2021-A05, and Tsinghua University Initiative Scientific Research Program (No. 20223080023). Z.Z. is supported by NSFC grants (No.11988101, 12041302, 11703036, and U1931110), CAS Interdisciplinary Innovation Team (JCTD-2019-05), and No.CMS-CSST-2021-A08. This project has received support from the European Union’s Horizon 2020 Research and Innovation Programme under the Marie Skłodowska-Curie grant agreement No 101086388. YJ.J is supported by the Cultivation Project for FAST Scientific Payoff and Research Achievement of CAMS-CAS. H.C. is supported by the Key Research Project of Zhejiang Lab (No. 2021PE0AC03). 
\end{acknowledgements}

\bibliographystyle{raa}
\bibliography{ref}{}

\begin{thebibliography}{67}
\providecommand\natexlab[1]{#1}
\providecommand\JournalTitle[1]{#1}

\bibitem[{Abazajian} {et~al.}(2009)]{Abazajian2009}
{Abazajian}, K.~N., {Adelman-McCarthy}, J.~K., {Ag{\"u}eros}, M.~A., {et~al.}
  2009, \apjs, 182, 543

\bibitem[{Alam} {et~al.}(2015)]{Alam2015}
{Alam}, S., {Albareti}, F.~D., {Allende Prieto}, C., {et~al.} 2015, \apjs, 219,
  12

\bibitem[{Bell} {et~al.}(2003)]{Bell2003}
{Bell}, E.~F., {McIntosh}, D.~H., {Katz}, N., \& {Weinberg}, M.~D. 2003, \apjs,
  149, 289

\bibitem[{Catinella} {et~al.}(2010)]{Catinella2010}
{Catinella}, B., {Schiminovich}, D., {Kauffmann}, G., {et~al.} 2010, \mnras,
  403, 683

\bibitem[{Catinella} {et~al.}(2012)]{Catinella2012}
{Catinella}, B., {Kauffmann}, G., {Schiminovich}, D., {et~al.} 2012, \mnras,
  420, 1959

\bibitem[{Colless} {et~al.}(2001)]{Colless2001}
{Colless}, M., {Dalton}, G., {Maddox}, S., {et~al.} 2001, \mnras, 328, 1039

\bibitem[{Cortese} {et~al.}(2021)]{Cortese2021}
{Cortese}, L., {Catinella}, B., \& {Smith}, R. 2021, \pasa, 38, e035

\bibitem[{D{\'e}nes} {et~al.}(2016)]{Denes2016}
{D{\'e}nes}, H., {Kilborn}, V.~A., {Koribalski}, B.~S., \& {Wong}, O.~I. 2016,
  \mnras, 455, 1294

\bibitem[{DESI Collaboration} {et~al.}(2023{\natexlab{a}})]{DESI2023b}
{DESI Collaboration}, {Adame}, A.~G., {Aguilar}, J., {et~al.}
  2023{\natexlab{a}}, arXiv e-prints, arXiv:2306.06308

\bibitem[{DESI Collaboration} {et~al.}(2023{\natexlab{b}})]{DESI2023}
{DESI Collaboration}, {Adame}, A.~G., {Aguilar}, J., {et~al.}
  2023{\natexlab{b}}, arXiv e-prints, arXiv:2306.06307

\bibitem[{Dey} {et~al.}(2019)]{Dey2019}
{Dey}, A., {Schlegel}, D.~J., {Lang}, D., {et~al.} 2019, \aj, 157, 168

\bibitem[{Driver} {et~al.}(2011)]{Driver2011}
{Driver}, S.~P., {Hill}, D.~T., {Kelvin}, L.~S., {et~al.} 2011, \mnras, 413,
  971

\bibitem[{Du} {et~al.}(2015)]{Du2015}
{Du}, W., {Wu}, H., {Lam}, M.~I., {et~al.} 2015, \aj, 149, 199

\bibitem[{Eke} {et~al.}(2006)]{Eke2006}
{Eke}, V.~R., {Baugh}, C.~M., {Cole}, S., {Frenk}, C.~S., \& {Navarro}, J.~F.
  2006, \mnras, 370, 1147

\bibitem[{Grogin} {et~al.}(2011)]{Grogin2011}
{Grogin}, N.~A., {Kocevski}, D.~D., {Faber}, S.~M., {et~al.} 2011, \apjs, 197,
  35

\bibitem[{Gunn} \& {Gott}(1972)]{Gunn1972}
{Gunn}, J.~E., \& {Gott}, J.~Richard, I. 1972, \apj, 176, 1

\bibitem[{Guo} {et~al.}(2013)]{Guo2013}
{Guo}, Q., {White}, S., {Angulo}, R.~E., {et~al.} 2013, \mnras, 428, 1351

\bibitem[{Guo} {et~al.}(2011)]{Guo2011}
{Guo}, Q., {White}, S., {Boylan-Kolchin}, M., {et~al.} 2011, \mnras, 413, 101

\bibitem[{Guo} {et~al.}(2020)]{Guo2020}
{Guo}, Q., {Hu}, H., {Zheng}, Z., {et~al.} 2020, Nature Astronomy, 4, 246

\bibitem[{Haynes} {et~al.}(1984)]{Haynes1984}
{Haynes}, M.~P., {Giovanelli}, R., \& {Chincarini}, G.~L. 1984, \araa, 22, 445

\bibitem[{Haynes} {et~al.}(2011)]{Haynes2011}
{Haynes}, M.~P., {Giovanelli}, R., {Martin}, A.~M., {et~al.} 2011, \aj, 142,
  170

\bibitem[{Haynes} {et~al.}(2018)]{Haynes2018}
{Haynes}, M.~P., {Giovanelli}, R., {Kent}, B.~R., {et~al.} 2018, \apj, 861, 49

\bibitem[{He} {et~al.}(2013)]{He2013}
{He}, Y.~Q., {Xia}, X.~Y., {Hao}, C.~N., {et~al.} 2013, \apj, 773, 37

\bibitem[{Hu} {et~al.}(2023)]{Hu2023}
{Hu}, H.-J., {Guo}, Q., {Zheng}, Z., {et~al.} 2023, \apjl, 947, L9

\bibitem[{Huchra} {et~al.}(2012)]{Huchra2012}
{Huchra}, J.~P., {Macri}, L.~M., {Masters}, K.~L., {et~al.} 2012, \apjs, 199,
  26

\bibitem[{Jiang} {et~al.}(2019)]{Jiang2019}
{Jiang}, P., {Yue}, Y., {Gan}, H., {et~al.} 2019, Science China Physics,
  Mechanics, and Astronomy, 62, 959502

\bibitem[{Jiang} {et~al.}(2020)]{Jiang2020}
{Jiang}, P., {Tang}, N.-Y., {Hou}, L.-G., {et~al.} 2020, Research in Astronomy
  and Astrophysics, 20, 064

\bibitem[{Jing} {et~al.}(2023)]{Jing2023}
{Jing}, Y., {Wang}, J., \& {FAST Collaboration}. 2023, In prep.

\bibitem[{Kannappan}(2004)]{Kannappan2004}
{Kannappan}, S.~J. 2004, \apjl, 611, L89

\bibitem[{Kennicutt}(1998)]{Kennicutt1998}
{Kennicutt}, Robert~C., J. 1998, \apj, 498, 541

\bibitem[{Kere{\v{s}}} {et~al.}(2005)]{Keres2005}
{Kere{\v{s}}}, D., {Katz}, N., {Weinberg}, D.~H., \& {Dav{\'e}}, R. 2005,
  \mnras, 363, 2

\bibitem[{Kroupa}(2002)]{Kroupa2002}
{Kroupa}, P. 2002, Science, 295, 82

\bibitem[{Krumholz} {et~al.}(2009)]{Krumholz2009}
{Krumholz}, M.~R., {McKee}, C.~F., \& {Tumlinson}, J. 2009, \apj, 693, 216

\bibitem[{Lawrence} {et~al.}(2007)]{Lawrence2007}
{Lawrence}, A., {Warren}, S.~J., {Almaini}, O., {et~al.} 2007, \mnras, 379,
  1599

\bibitem[{Leroy} {et~al.}(2008)]{Leroy2008}
{Leroy}, A.~K., {Walter}, F., {Brinks}, E., {et~al.} 2008, \aj, 136, 2782

\bibitem[{Li} {et~al.}(2012)]{Li2012}
{Li}, C., {Kauffmann}, G., {Fu}, J., {et~al.} 2012, \mnras, 424, 1471

\bibitem[{Lisker} {et~al.}(2007)]{Lisker2007}
{Lisker}, T., {Grebel}, E.~K., {Binggeli}, B., \& {Glatt}, K. 2007, \apj, 660,
  1186

\bibitem[{Liu} {et~al.}(2008)]{Liu2008}
{Liu}, F.~S., {Xia}, X.~Y., {Mao}, S., {Wu}, H., \& {Deng}, Z.~G. 2008, \mnras,
  385, 23

\bibitem[{Madau} \& {Dickinson}(2014)]{Madau2014}
{Madau}, P., \& {Dickinson}, M. 2014, \araa, 52, 415

\bibitem[{Martin} {et~al.}(2005)]{Martin2005}
{Martin}, D.~C., {Fanson}, J., {Schiminovich}, D., {et~al.} 2005, \apjl, 619,
  L1

\bibitem[{McKee} \& {Krumholz}(2010)]{McKee2010}
{McKee}, C.~F., \& {Krumholz}, M.~R. 2010, \apj, 709, 308

\bibitem[{McNichols} {et~al.}(2016)]{McNichols2016}
{McNichols}, A.~T., {Teich}, Y.~G., {Nims}, E., {et~al.} 2016, \apj, 832, 89

\bibitem[{Moore} {et~al.}(1996)]{Moore1996}
{Moore}, B., {Katz}, N., {Lake}, G., {Dressler}, A., \& {Oemler}, A. 1996,
  \nat, 379, 613

\bibitem[{Nan}(2006)]{Nan2006}
{Nan}, R. 2006, Science in China: Physics, Mechanics and Astronomy, 49, 129

\bibitem[{Nelson} {et~al.}(2019)]{Nelson2019}
{Nelson}, D., {Pillepich}, A., {Springel}, V., {et~al.} 2019, \mnras, 490, 3234

\bibitem[{Oh} {et~al.}(2015)]{Oh2015}
{Oh}, S.-H., {Hunter}, D.~A., {Brinks}, E., {et~al.} 2015, \aj, 149, 180

\bibitem[{Paturel} {et~al.}(2003)]{Paturel2003}
{Paturel}, G., {Theureau}, G., {Bottinelli}, L., {et~al.} 2003, \aap, 412, 57

\bibitem[{Pillepich} {et~al.}(2019)]{Pillepich2019}
{Pillepich}, A., {Nelson}, D., {Springel}, V., {et~al.} 2019, \mnras, 490, 3196

\bibitem[{Rhee} {et~al.}(2023)]{Rhee2023}
{Rhee}, J., {Meyer}, M., {Popping}, A., {et~al.} 2023, \mnras, 518, 4646

\bibitem[{Roberts}(1962)]{Roberts1962}
{Roberts}, M.~S. 1962, \aj, 67, 437

\bibitem[{Robotham} {et~al.}(2011)]{Robotham2011}
{Robotham}, A.~S.~G., {Norberg}, P., {Driver}, S.~P., {et~al.} 2011, \mnras,
  416, 2640

\bibitem[{Salim} {et~al.}(2007)]{Salim2007}
{Salim}, S., {Rich}, R.~M., {Charlot}, S., {et~al.} 2007, \apjs, 173, 267

\bibitem[{Saulder} {et~al.}(2016)]{Saulder2016}
{Saulder}, C., {van Kampen}, E., {Chilingarian}, I.~V., {Mieske}, S., \&
  {Zeilinger}, W.~W. 2016, \aap, 596, A14

\bibitem[{Scoville} {et~al.}(2007)]{Scoville2007}
{Scoville}, N., {Aussel}, H., {Brusa}, M., {et~al.} 2007, \apjs, 172, 1

\bibitem[{Skrutskie} {et~al.}(2006)]{Skrutskie2006}
{Skrutskie}, M.~F., {Cutri}, R.~M., {Stiening}, R., {et~al.} 2006, \aj, 131,
  1163

\bibitem[{Springel} \& {Hernquist}(2003)]{Springel2003}
{Springel}, V., \& {Hernquist}, L. 2003, \mnras, 339, 289

\bibitem[{Springel} {et~al.}(2005)]{Springel2005}
{Springel}, V., {White}, S. D.~M., {Jenkins}, A., {et~al.} 2005, \nat, 435, 629

\bibitem[{Springob} {et~al.}(2005)]{Springob2005}
{Springob}, C.~M., {Haynes}, M.~P., {Giovanelli}, R., \& {Kent}, B.~R. 2005,
  \apjs, 160, 149

\bibitem[{White} \& {Rees}(1978)]{White1978}
{White}, S.~D.~M., \& {Rees}, M.~J. 1978, \mnras, 183, 341

\bibitem[{Wright} {et~al.}(2010)]{Wright2010}
{Wright}, E.~L., {Eisenhardt}, P. R.~M., {Mainzer}, A.~K., {et~al.} 2010, \aj,
  140, 1868

\bibitem[{Yang} {et~al.}(2007)]{Yang2007}
{Yang}, X., {Mo}, H.~J., {van den Bosch}, F.~C., {et~al.} 2007, \apj, 671, 153

\bibitem[{York} {et~al.}(2000)]{York2000}
{York}, D.~G., {Adelman}, J., {Anderson}, John~E., J., {et~al.} 2000, \aj, 120,
  1579

\bibitem[{Zhang} {et~al.}(2021{\natexlab{a}})]{Zhang2021}
{Zhang}, K., {Wu}, J., {Li}, D., {et~al.} 2021{\natexlab{a}}, \mnras, 500, 1741

\bibitem[{Zhang} {et~al.}(2021{\natexlab{b}})]{Zhangw2021}
{Zhang}, W., {Kauffmann}, G., {Wang}, J., {et~al.} 2021{\natexlab{b}}, \aap,
  648, A25

\bibitem[{Zhang} {et~al.}(2009)]{Zhang2009}
{Zhang}, W., {Li}, C., {Kauffmann}, G., {et~al.} 2009, \mnras, 397, 1243

\bibitem[{Zheng} {et~al.}(2020)]{Zheng2020}
{Zheng}, Z., {Li}, D., {Sadler}, E.~M., {Allison}, J.~R., \& {Tang}, N. 2020,
  \mnras, 499, 3085

\bibitem[{Zheng} {et~al.}(2015)]{Zheng2015}
{Zheng}, Z., {Thilker}, D.~A., {Heckman}, T.~M., {et~al.} 2015, \apj, 800, 120

\end{thebibliography}


\end{document}